\documentclass[conference]{IEEEtran}

\usepackage{amsmath,amssymb}
\usepackage{booktabs}
\usepackage{array}
\usepackage{cite}
\usepackage{url}
\usepackage{pgfplots}
\pgfplotsset{compat=1.18}

\title{Predictive Traffic Shaping as a UE--Network\\
Control Loop in Wireless Systems}

\author{
\IEEEauthorblockN{Subramanian Vasudevan}
\IEEEauthorblockA{Independent Researcher\\
Morristown, NJ, USA}
\and
\IEEEauthorblockN{Shriram Vasudevan}
\IEEEauthorblockA{Independent Researcher\\
Morristown, NJ, USA}
}

\begin{document}

\maketitle

\begin{abstract}
Wireless systems usually react to current channel conditions, queue state, and policy. Yet service conditions can often be anticipated seconds ahead. This paper studies predictive traffic shaping, a slower user-equipment (UE) control loop that changes when flexible demand reaches the radio access network (RAN). The UE estimates useful pre-event demand and releases it across a lookahead window. Cooperative deployments may also send a compact future-risk descriptor to the network. The trigger uses prediction confidence. Predictive service is confined to available surplus, while a debt account preserves long-term fairness after temporary pre-event preference. A bandwidth-time model captures efficiency gains and load smoothing. It also accounts for prediction waste and shared-resource cost. In a stylized shared-cell simulation, paced demand release substantially expands the stable-feasible region. A safe service cap and admission control provide further gains at longer windows. PRISM, an application-owned middleware prototype, implements the local control policy using ordinary mobile transfer mechanisms.
\end{abstract}


\section{Introduction}

Wireless control is predominantly reactive. Applications generate traffic. The UE exposes queues and channel reports. The RAN then allocates resources according to current conditions. This structure supports millisecond scheduling, while slower service changes may be predictable seconds in advance.

Recurring geometry, blockage, mobility, and load create many of these changes. High-frequency deployments are especially sensitive to blockage and spatial propagation conditions~\cite{rappaport2013}. Many mobile transfers also have timing flexibility. Video segments and map tiles can often be sent early. Model updates, telemetry, and background synchronization can often wait. Modern devices expose motion and location signals that can serve as inputs to lightweight service forecasts~\cite{androidsensors,applemotion,applelocation}. Route history and measured bandwidth can support prediction of upcoming service conditions~\cite{borkowski2016,riiser2012,sethuraman2021}. Relevant inputs may also include serving-cell history, recent throughput, and application class. Prior work covers anticipatory networking~\cite{bui2017}, mobile prefetching~\cite{higgins2012,borkowski2016}, predictive video delivery and adaptation~\cite{riiser2012,sun2016}, device-side bandwidth forecasts such as Foresight~\cite{sethuraman2021}, predictive network scheduling~\cite{bang2008,yu2016}, and Access Traffic Steering, Switching, and Splitting (ATSSS)~\cite{atsss2026}. We study a related control question: how should a UE use a service forecast to change the arrival process presented to the network? Shared capacity makes this question distinct. When many UEs advance traffic into the same interval, their actions can create a new demand peak.

Predictive traffic shaping is modeled as a slow UE-side loop that operates alongside fast RAN scheduling. The UE identifies an upcoming service event. It classifies traffic by temporal flexibility and estimates useful pre-event demand. It then changes release times before the event. In cooperative mode, the UE sends a compact future-risk descriptor. The network retains control of admission and resource allocation.

The paper makes one architectural contribution: a stabilized UE-side demand-shaping loop for shared wireless resources. Four parts develop this contribution:
\begin{enumerate}
    \item a future-risk descriptor for scheduling-relevant event information;
    \item a shaping policy with paced release, safe-surplus service, admission, and fairness accounting;
    \item a bandwidth-time model and shared-cell simulation; and
    \item PRISM, an app-owned middleware prototype for the local control policy.
\end{enumerate}
Service prediction is an input to the loop. A segmented map or an existing bandwidth forecaster may provide the event estimate. A lightweight learned model may provide the same fields. The controller uses event timing, expected loss, confidence, and duration.

\begin{table}[t]
\caption{Control Scope of Representative Predictive Wireless Systems}
\label{tab:scope}
\centering
\footnotesize
\begin{tabular}{p{0.22\columnwidth}p{0.25\columnwidth}p{0.43\columnwidth}}
\toprule
Approach & Control point & Shared-resource mechanism \\
\midrule
Application prefetching and adaptation~\cite{higgins2012,borkowski2016,riiser2012,sun2016}
& Fetch and bitrate timing
& Buffering, prediction, and bitrate control \\
Foresight~\cite{sethuraman2021}
& Device bandwidth plans
& Device bandwidth manager across applications \\
Predictive scheduling~\cite{bang2008,yu2016}
& RAN resource allocation
& Scheduler optimization using future channel state \\
ATSSS~\cite{atsss2026}
& Access-path selection
& UE--network steering, switching, and splitting \\
This work
& Timing of aggregate UE demand
& Pacing, surplus caps, admission, and fairness debt \\
\bottomrule
\end{tabular}
\end{table}

\section{Positioning Relative to Prior Work}

Related systems place prediction at different control points. Application prefetchers and predictive video systems schedule content or adapt bitrate for one application. Foresight exposes future bandwidth plans to multiple applications. Predictive schedulers use forecasts within the RAN. ATSSS coordinates traffic across access paths. Predictive traffic shaping controls the timing of aggregate UE demand and coordinates concurrent actions through shared-resource safeguards.

Foresight is the closest device-side precedent. It provides route-based bandwidth forecasts and allocates schedules across applications~\cite{sethuraman2021}. Predictive traffic shaping uses a similar forecast to control the aggregate arrival process. It also distinguishes traffic by temporal flexibility.

Network-side predictive scheduling operates at a different control point. It adjusts allocation after demand reaches the scheduler. The proposed loop adjusts release timing at the UE. The UE acts over hundreds of milliseconds to seconds, while the RAN remains the fast arbiter of shared resources. ATSSS provides a precedent for UE--network policy coordination~\cite{atsss2026}. Its control variable is the selected access path.

Energy-aware systems also exploit transfer timing~\cite{balasubramanian2009,schulman2010}. Some delay or batch transfers. Others favor stronger-signal route segments. Predictive shaping adds explicit control of the externality among UEs. Pacing, surplus caps, admission, and fairness accounting coordinate their shared use of favorable intervals.

\section{System Model and Future-Risk Descriptor}

Time is divided into slots $t=1,\ldots,T$ of duration $\Delta$. UE $i$ experiences an effective service process $C_{i,t}$. We measure it in useful bits per slot or application-weighted service units. Effective service reflects radio rate and scheduling opportunity. It may also reflect competing load, access technology, and application usefulness. The predictor supplies
\begin{equation}
\widehat{P}_{i,t}(c)=\Pr(C_{i,t}=c\mid x_{i,t}),
\end{equation}
where $x_{i,t}$ is the local context. It may contain mobility state or serving-cell history. Recent throughput and application class provide additional context. The controller summarizes the forecast as an event time, duration, expected loss, and confidence.

Applications expose different amounts of temporal flexibility. Advanceable traffic can move earlier. Video segments and map tiles are common examples. Deferrable traffic can wait for a lower-cost interval. Non-shiftable traffic requires immediate service. Suppressible traffic may be dropped when its expected utility is low.

Let $b_i^u$ be UE $i$'s useful pre-event demand cap. It measures the amount of traffic that remains useful when advanced. For the evaluation in Section~\ref{sec:simulation}, $b_i^u$ is instantiated as the deadzone content requirement $K$. Advancement stops when this cap is met.

In cooperative mode, the UE may report
\begin{equation}
D_i=(\tau_i,T_i,\ell_i,b_i^u,\sigma_i,\kappa_i,d_i),
\end{equation}
where $\tau_i$ is time-to-event and $T_i$ is expected duration. The expected service loss is $\ell_i$, and $\sigma_i$ is confidence. The traffic class is $\kappa_i$. The term $d_i$ records fairness debt or recent priority history. The UE keeps raw location, route, sensor streams, and application payloads local.

A practical predictor can begin with a segmented service map. The map may store degradation probability and expected duration. Mean service and variance characterize the event further. A lightweight learned model can smooth across nearby states. Map- or digital-twin-derived features can support richer deployments. These choices affect forecast quality. Shared-resource safeguards remain necessary when many UEs act on related forecasts.

\section{Predictive UE--Network Control Loop}

The loop runs below individual applications and above the radio scheduler. At the application layer, it can use rate-limited transfer scheduling, HTTP range requests, and background-transfer APIs. OS-level deployments can add traffic classes, transport-level pacing, and multipath policy. This slower controller updates at epochs well above the RAN scheduler's slot-level timescale.

At each control epoch, the UE first identifies a future service event. It then classifies candidate traffic and estimates useful pre-event demand. The controller evaluates expected gain and assigns a release schedule. After the useful cap is reached, it clears temporary priority and updates fairness debt.

In UE-only mode, the device controls the timing and rate of its own traffic. In cooperative RAN mode, it sends (2). The network can admit predictive demand and apply a bounded scheduler adjustment. One possible mapping is
\begin{equation}
w_i(t)=w_i^0(t)\bigl(1+\gamma q_i(t)\bigr),
\quad 0\leq\gamma q_i(t)\leq\Gamma,
\end{equation}
where $w_i^0(t)$ is the baseline weight and $\Gamma$ caps the adjustment. Equation (3) gives a bounded mapping from future risk to scheduling weight. The adjustment expires with the descriptor or when the useful bytes are served. Its resource use is charged to a surplus or fairness budget.

Per-application queues may carry deadline slack and utility. They may also record traffic class and maximum useful advance. The network receives the aggregate descriptor.

\section{Predictive Control Policy}

\subsection{Bandwidth-Time Objective}

Bandwidth-time is the relevant resource measure because the same transfer can consume different amounts of airtime or spectrum under different channel and load conditions. A transfer performed during periods of high spectral efficiency or low network load generally consumes fewer radio resources than the same transfer performed during congestion or poor radio conditions.

Let $\eta_{i,t}$ denote the spectral efficiency experienced by UE $i$ at time $t$, and let $\lambda_t$ denote the marginal value, or scarcity price, of one unit of bandwidth-time during slot $t$. Ignoring fixed conversion constants, the resource cost of transmitting one bit is
\begin{equation}
\phi_{i,t}=\frac{\lambda_t}{\eta_{i,t}}.
\end{equation}
Equation (4) states that the same data transfer becomes less expensive whenever spectral efficiency increases or the network becomes less congested.

Predictive transmission is worthwhile only when the reduction in bandwidth-time cost outweighs the penalties introduced by moving traffic earlier. Let $G_{\mathrm{SE},i}$ denote the resource savings obtained from transmitting during periods of higher spectral efficiency, and let $G_{\mathrm{load},i}$ denote the benefit obtained by smoothing network load. Let $P_{\mathrm{inflation},i}$ denote additional traffic induced by predictive transmission, and let $P_{\mathrm{externality},i}$ denote the cost imposed on competing users through additional resource consumption. Define the net value realized when the predicted event occurs as
\begin{equation}
V_i=G_{\mathrm{SE},i}+G_{\mathrm{load},i}
-P_{\mathrm{inflation},i}-P_{\mathrm{externality},i}.
\end{equation}
The penalty $P_{\mathrm{waste},i}$ is reserved for the incorrect-prediction branch in Section~V-C; it represents the total cost of traffic advanced unnecessarily, including the corresponding resource use.

Congestion effects are represented through the ordinary offered load $D_t$, predictive load $x_t$, and total service capacity $A_t$. A convex congestion cost
\begin{equation}
H_t(D_t+x_t)=\psi\!\left(\frac{D_t+x_t}{A_t}\right),
\qquad \psi''(\cdot)\geq 0,
\end{equation}
captures the fact that congestion increases disproportionately as utilization approaches capacity. Moving demand away from heavily loaded intervals therefore yields greater benefit than adding the same demand to lightly loaded intervals. Conversely, if many UEs simultaneously advance traffic into the same interval, the resulting congestion reduces the benefit of prediction.

Equations (4)--(6) define the design objective rather than an optimization solved online. The controller does not explicitly estimate every term. Instead, the triggering, pacing, and admission policies developed below are computationally simple approximations that pursue the same qualitative objectives: transmitting during low-cost intervals, exploiting surplus bandwidth-time, reducing congestion peaks, and limiting shared-resource externalities.

\subsection{Prediction-Critical Feasibility}

Prediction is only valuable when advance transmission changes whether an application's deadline can be met.

Suppose an application requires an object of size $B$ bits before or during a predicted service degradation. Let $C_G$ denote the useful service available during the degradation itself, and let $C_P$ denote the useful service that can be accumulated beforehand using surplus capacity. Reactive service is feasible only if
\begin{equation}
B\leq C_G,
\end{equation}
whereas predictive service becomes feasible whenever
\begin{equation}
B\leq C_P+C_G.
\end{equation}
The interval
\begin{equation}
C_G<B\leq C_P+C_G
\end{equation}
defines the prediction-critical region. Within this region, the application cannot meet its deadline through reactive transmission alone, but succeeds if sufficient data are preloaded before the degradation begins. Prediction therefore enlarges the feasible operating region by exploiting otherwise unused surplus capacity.

\subsection{Uncertainty and Triggering}

Prediction should influence scheduling only when its expected benefit remains positive despite uncertainty. Let $\sigma_i\in[0,1]$ denote the calibrated probability that UE $i$'s predicted event occurs. Using the correct-prediction value in (5), the expected value of advancing traffic is
\begin{equation}
E_i=\sigma_iV_i-(1-\sigma_i)P_{\mathrm{waste},i}.
\end{equation}
Equation (10) is the uncertainty-aware version of the objective in Section~V-A. If the event occurs, predictive transmission realizes the efficiency and load-smoothing value represented by $V_i$. If the prediction is wrong, the controller incurs $P_{\mathrm{waste},i}$ because traffic was transmitted unnecessarily.

The controller advances predictive traffic whenever $E_i>0$. Rather than having all UEs act simultaneously, a randomized trigger converts expected value into an advancement probability,
\begin{equation}
\pi_i=1-\exp\!\left(-\alpha[E_i]_+\right),
\end{equation}
where $[z]_+=\max\{z,0\}$ and $\alpha$ scales the response. As expected benefit increases, the probability of advancing approaches one. For non-positive expected value, the trigger remains inactive. Randomization reduces synchronization among UEs that observe similar predictions.

\subsection{Pacing, Safe Serving, and Admission}

Triggering determines whether predictive transmission should occur. The remaining mechanisms determine how predictive traffic is released and how much can be safely served.

Let $S_t$ denote the surplus bandwidth-time remaining after ordinary traffic has been served during slot $t$. We call the following bound the safe service cap: it limits predictive traffic to the estimated capacity that remains after ordinary application traffic has been protected. Predictive traffic is limited to a fraction $\rho$ of this surplus,
\begin{equation}
\sum_i\frac{y_{i,t}}{\eta_{i,t}}\leq\rho S_t,
\qquad 0<\rho<1,
\end{equation}
where $y_{i,t}$ denotes predictive traffic served for UE $i$ during slot $t$. Dividing by spectral efficiency converts transmitted bits into consumed bandwidth-time. A UE-only implementation estimates this surplus locally, whereas a cooperative RAN implementation can enforce it directly.

Once a UE has decided to advance traffic, that traffic should be released gradually rather than in a burst. Let $r_i(W_R)$ denote the total bandwidth-time required to advance UE $i$'s useful demand over a release window of duration $W_R$, and let $R_t^{\mathrm{off}}$ denote aggregate predictive bandwidth-time offered at time $t$. Every release schedule obeys the average-load lower bound
\begin{equation}
\max_t R_t^{\mathrm{off}}
\geq \frac{1}{W_R}\sum_{i=1}^{N}r_i(W_R).
\end{equation}
Flat pacing attains this lower bound when demand is divisible and the available rate is constant. Randomized release offsets further reduce synchronization while remaining close to the minimum average offered load.

Pacing controls offered predictive load. A separate admission mechanism controls served predictive load. When ordinary traffic consumes the available surplus, predictive backlog may accumulate. Stabilization therefore combines paced release with surplus-limited service and admission control.

Let $\mathcal{W}_R$ denote the slots in the release window. The aggregate admission fraction is
\begin{equation}
a_{\mathrm{adm}}=\min\!\left\{1,
\frac{\sum_{t\in\mathcal{W}_R}\rho S_t}
{\sum_i r_i(W_R)}\right\},
\end{equation}
for positive predictive demand. Equation (14) compares predictive capacity available during the release window with total requested predictive demand. When sufficient surplus exists, all predictive traffic is admitted. Otherwise, only the corresponding fraction can be served. The scheduler may convert this aggregate fraction into per-UE admission decisions using confidence, traffic class, or fairness state.

Finally, once a UE has received its useful predictive service, any temporary priority should disappear. Let $B_i^{\mathrm{served}}(t)$ denote cumulative predictive bytes served through slot $t$. For $b_i^u>0$, the remaining predictive priority decays linearly,
\begin{equation}
q_i(t)=q_i(0)\left[1-
\frac{B_i^{\mathrm{served}}(t)}{b_i^u}\right]_+.
\end{equation}
Priority therefore decreases continuously as useful demand is satisfied, reaching zero once the useful-demand cap has been met. Predictive service is also recorded as fairness debt, which the proposed controller uses to offset temporary pre-event priority over time. This accounting follows the broader goal of preserving long-term fairness in shared network allocation~\cite{kelly1998}.

\section{Simulation Study}
\label{sec:simulation}

\subsection{Scope, Scenario, and Metrics}

The simulation isolates the relationship between lookahead, pacing, and shared-cell surplus. It uses a stylized single-cell resource model. The main output is the stable-feasible region as the release window and number of UEs vary.

We simulate one slotted downlink cell with $\Delta t=0.25$~s and 100~MHz bandwidth. Ordinary traffic is served first and uses about 58\% of capacity on average, with sinusoidal variation and random noise. The remainder is surplus $S_t$. Sending $x$ predictive bits to UE $i$ consumes $x/\eta_{i,t}$ MHz$\cdot$s. Predictive service is restricted to $\rho=0.65$ of surplus. Under the nominal load, total surplus is about 10.5~MHz$\cdot$s per slot and the safe envelope is about 6.8~MHz$\cdot$s.

Each run creates $N$ UEs approaching the same degradation interval, representing a tunnel, blockage zone, cell edge, or crowded venue. The gap lasts $G\approx8$~s. Spectral efficiency is $\eta_{\mathrm{pre}}\approx3$ bit/s/Hz before the gap and $\eta_{\mathrm{gap}}\approx0.12$ bit/s/Hz during it. A 3~Mbps stream needs $K=24$~Mbits for an eight-second gap. That demand costs about 8~MHz$\cdot$s before the gap and 200~MHz$\cdot$s during it.

A run is stable-feasible when at least 95\% of UEs receive at least 95\% of useful demand before the gap ends. Predictive service must also exceed the safe envelope in no more than 5\% of slots. The boundary $N_{\max}(W_R)$ is the largest tested $N$ for which at least 80\% of 48 random seeds are stable-feasible. A 90\% seed threshold is used as a sensitivity check. We test $W_R\in\{5,10,15,20,30,45,60,90\}$~s.

\begin{table}[t]
\caption{Mechanism Ladder Evaluated in Simulation}
\label{tab:policies}
\centering
\footnotesize
\begin{tabular}{lcccc}
\toprule
Policy & Lookahead & Pacing & Safe cap & Admission \\
\midrule
Reactive   & -- & -- & -- & -- \\
Naive      & $\checkmark$ & -- & -- & -- \\
Paced      & $\checkmark$ & $\checkmark$ & -- & -- \\
Stabilized & $\checkmark$ & $\checkmark$ & $\checkmark$ & $\checkmark$ \\
\bottomrule
\end{tabular}
\end{table}

\subsection{Policies}

Table~\ref{tab:policies} lists the four policies. Reactive releases demand at the start of the gap. Naive predictive releases all eligible demand at the start of the lookahead window. Paced predictive releases each UE's demand at an approximately constant rate over $W_R$. Stabilized predictive adds the safe service cap, useful-byte cap, and admission when total demand exceeds integrated safe surplus.

Reactive service is limited by low gap efficiency. Naive prediction can deliver many bits before the gap but presents a synchronized burst. Pacing lowers the offered bandwidth-time rate toward
\begin{equation}
R^{\mathrm{off}}\approx\frac{NK}{\eta_{\mathrm{pre}}W_R}.
\end{equation}
Near the feasibility boundary, ordinary-load fluctuations can create backlog and later served-load spikes. The stabilized policy clips those spikes and rejects demand above the safe budget.

\subsection{Results}

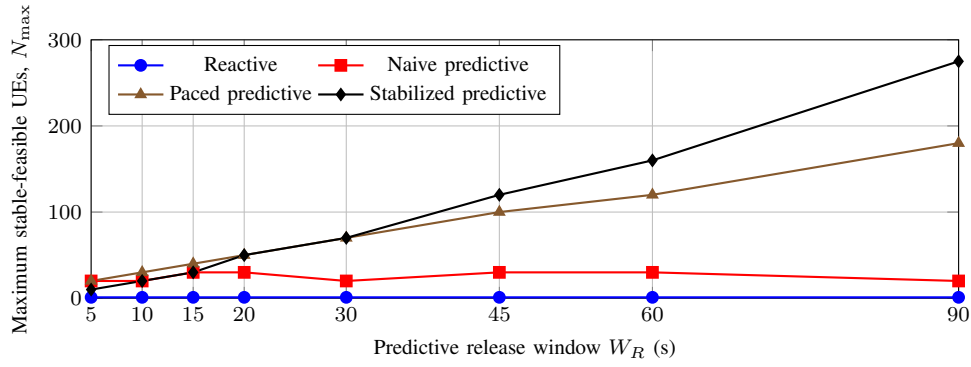
\begin{figure*}[t]
\centering
\begin{tikzpicture}
\begin{axis}[
    width=0.72\textwidth,
    height=5.0cm,
    xlabel={Predictive release window $W_R$ (s)},
    ylabel={Maximum stable-feasible UEs, $N_{\max}$},
    xmin=5,xmax=90,
    ymin=0,ymax=300,
    xtick={5,10,15,20,30,45,60,90},
    ytick={0,100,200,300},
    grid=major,
    legend columns=2,
    legend style={at={(0.02,0.98)},anchor=north west,font=\footnotesize},
    tick label style={font=\footnotesize},
    label style={font=\footnotesize}
]
\addplot[blue,mark=*,thick] coordinates {(5,1)(10,1)(15,1)(20,1)(30,1)(45,1)(60,1)(90,1)};
\addlegendentry{Reactive}
\addplot[red,mark=square*,thick] coordinates {(5,20)(10,20)(15,30)(20,30)(30,20)(45,30)(60,30)(90,20)};
\addlegendentry{Naive predictive}
\addplot[brown!70!black,mark=triangle*,thick] coordinates {(5,20)(10,30)(15,40)(20,50)(30,70)(45,100)(60,120)(90,180)};
\addlegendentry{Paced predictive}
\addplot[black,mark=diamond*,thick] coordinates {(5,10)(10,20)(15,30)(20,50)(30,70)(45,120)(60,160)(90,275)};
\addlegendentry{Stabilized predictive}
\end{axis}
\end{tikzpicture}
\caption{Stable-feasible boundary under shared-cell constraints. Naive prediction remains safety-limited. Pacing produces the main expansion, and safe-cap service with admission adds capacity at moderate and long windows.}
\label{fig:boundary}
\end{figure*}

Table~\ref{tab:boundary} and Fig.~\ref{fig:boundary} report the stable-feasible boundary. Reactive service remains at one UE for every release window. Naive prediction supports roughly 20--30 UEs and remains nearly flat. A longer window moves the same synchronized release earlier while preserving the burst.

The paced boundary rises with $W_R$ and reaches 180 UEs at 90~s. This gain comes from spreading traffic across the release window. Stabilized shaping matches pacing at 20 and 30~s and supports 120, 160, and 275 UEs at 45, 60, and 90~s, compared with 100, 120, and 180 for pacing. The 90~s gain is 53\% relative to pacing.

At 5 and 10~s, aggregate demand exceeds integrated safe surplus, so the stabilized policy rejects some UEs. Paced mode attempts to serve those UEs and has a higher delivery boundary under the chosen violation threshold. At longer windows, the cap prevents backlog catch-up from producing served-load violations. The stricter 90\% seed threshold preserves the same broad ordering.

The results show that pre-gap service can make a deadline feasible when gap capacity is insufficient. Unsmooth predictive release relocates congestion. Pacing limits offered load. Safe service and admission keep served load within the modeled envelope.

This experiment evaluates stabilization at the resource-allocation level. Possible extensions include measuring the incremental value of descriptor exchange and integrating the policy with production-oriented RAN schedulers.

\begin{table}[t]
\caption{Stable-Feasible Boundary $N_{\max}(W_R)$ at the 80\% Seed-Feasibility Threshold}
\label{tab:boundary}
\centering
\footnotesize
\setlength{\tabcolsep}{3.1pt}
\begin{tabular}{lrrrrrrrr}
\toprule
$W_R$ (s) & 5 & 10 & 15 & 20 & 30 & 45 & 60 & 90 \\
\midrule
Reactive   & 1  & 1  & 1  & 1  & 1  & 1   & 1   & 1 \\
Naive      & 20 & 20 & 30 & 30 & 20 & 30  & 30  & 20 \\
Paced      & 20 & 30 & 40 & 50 & 70 & 100 & 120 & 180 \\
Stabilized & 10 & 20 & 30 & 50 & 70 & 120 & 160 & 275 \\
\bottomrule
\end{tabular}
\end{table}

\subsection{Interpretation and Limitations}

The reported values of $N_{\max}$ should not be read as estimates of commercial cell capacity. They measure the number of simultaneous predictive flows that satisfy the delivery and safety criteria in the specific resource model above. The useful comparison is therefore across policies under the same load trace. In that comparison, naive lookahead changes the time of the burst but not its structure; pacing lowers offered-load concentration; and the safe cap with admission prevents accumulated backlog from being served in a later spike.

The scenario intentionally places all UEs near the same degradation interval. This is a demanding case because event times are correlated and release windows overlap. More heterogeneous event times would usually spread predictive demand, whereas correlated forecast error could waste surplus or admit traffic for an event that does not occur. The model also abstracts from handover, retransmissions, per-flow transport dynamics, and control-message overhead. Ordinary load is synthetic, and spectral efficiency is represented by pre-gap and in-gap regimes rather than a full channel trace.

These choices isolate the shared-resource question but limit the interpretation of absolute values. A deployment study should replay measured mobility and load traces, include calibrated event probabilities, and evaluate end-to-end outcomes such as stall time, completion latency, predictive waste, and ordinary-user throughput. The present result is narrower: under a common stress case, lookahead is useful only when release and service are stabilized.

\section{PRISM Middleware Prototype and Deployment Path}

PRISM is an application-owned UE-side middleware that shapes an app's transfers before a predicted connectivity gap. It operates within a single application. System-wide shaping requires operating-system privileges, enterprise management, or special entitlements. The prototype evaluates the local policy logic.

The implementation consists of a deterministic Swift policy package and a SwiftUI demonstration app that issues ordinary \texttt{URLSession} transfers to a controllable segment server. A service-event module constructs the fields in (2). In UE-only mode, these fields remain internal to PRISM; only a cooperative deployment transmits the descriptor to the network. A demand ledger records useful bytes remaining, bytes delivered before the event, and fairness debt. The policy engine selects advance, defer, suppress, or normal service; the transfer layer executes the result.

The policy engine first checks whether an event is active and whether the traffic is shiftable. It then checks confidence, remaining useful demand, and modeled gain. Eligible actions are limited by a local safe-service estimate:
\begin{equation}
B_{\mathrm{safe}}=\rho S_{\mathrm{app}}W_{\mathrm{eff}},
\qquad
r_{\mathrm{rel}}=
\frac{\min\{b^u_{\mathrm{rem}},B_{\max},B_{\mathrm{safe}}\}}
{W_{\mathrm{eff}}}.
\end{equation}
Here, $S_{\mathrm{app}}$ is a conservative estimate of surplus application throughput in bytes per second. The capped release window is $W_{\mathrm{eff}}$. The configured per-event byte limit is $B_{\max}$. The ledger prevents duplicate fetching and stops advancement when the cap is exhausted.

The same policy supports three deployment levels. The app-owned level is available today and controls one application's transfers. An operating-system or enterprise level could coordinate queues across the device. A cooperative RAN level could accept the descriptor and enforce shared admission. It could also enforce surplus caps, expiry, and fairness accounting. PRISM establishes the implementability of the local decisions. The shared-cell simulation evaluates the value of network-side enforcement when many UEs act together.

\subsection{Fallback Behavior}

The app-owned controller reverts to ordinary transfer behavior when no event is active, the descriptor has expired, confidence falls below the trigger, or no conservative surplus estimate is available. An event update may reduce the remaining release rate, but it does not create new application demand; only bytes already marked as useful and shiftable are eligible. If an advanced object has already been fetched, the ledger records it as satisfied so that a later event update does not fetch it again. These rules make prediction advisory rather than mandatory. A wrong or stale forecast can consume some surplus and energy, but it does not block non-shiftable traffic or leave the application dependent on continued prediction. Cross-application guarantees still require operating-system or RAN support, because an app-owned controller cannot observe competing queues or enforce a cell-wide surplus cap.

\section{Discussion}

Predictive shaping can increase traffic before an event. Its value comes from moving useful bytes to intervals with lower bandwidth-time cost. The release policy must also avoid a new peak. Useful demand delivered and radio-resource cost are therefore central outputs. Predictive waste and safety-envelope violations provide complementary measures.

Cooperative signaling introduces trust requirements. Descriptors should be authenticated and rate-limited. The scheduler should cap their influence and maintain an audit trail. Repeatedly inaccurate reports could reduce admission weight or increase fairness debt.

Privacy is addressed through descriptor scope. The report carries event timing, duration, useful demand, confidence, traffic class, and priority history. Raw route and application data remain local to the UE.

Traffic elasticity also determines the opportunity for shaping. Streaming and background transfers may provide substantial useful demand. Interactive traffic may provide little. UE-only deployments also depend on conservative estimates of available surplus.

\section{Conclusion}

Predictive traffic shaping lets a UE decide when flexible demand should enter the network. The loop identifies useful traffic and distributes advance transfers across the available window. Shared surplus bounds predictive service. Temporary pre-event preference is tracked and later offset through fairness accounting. A compact future-risk descriptor can support network-side admission and scheduling.

The shared-cell simulation shows that pacing expands the feasible region. Safe-cap service and admission add further capacity at longer windows. PRISM implements the local policy with existing application mechanisms.

Possible directions include evaluation under realistic mobility and calibrated prediction error; extension to multicell operation, physical-layer retransmissions, and handover; integration with production-oriented RAN schedulers; and measurement of descriptor overhead, operator trust, and adversarial reporting in cooperative deployments.

\section*{Acknowledgment}

OpenAI ChatGPT was used to assist with language editing and organization during manuscript preparation. The authors reviewed and verified all technical content, equations, references, claims, and conclusions.

\end{document}